\documentclass[aps, twocolumn, superscriptaddress]{revtex4-2} 

\usepackage[switch]{lineno}
\usepackage{amsmath}
\usepackage{amssymb}
\usepackage{empheq}
\usepackage[parfill]{parskip}
\usepackage[active]{srcltx}
\usepackage{color}
\usepackage{array}
\usepackage{booktabs}
\usepackage{amsfonts}
\usepackage{dsfont}
\usepackage{graphicx}
\usepackage{appendix}
\usepackage{hyperref}
\usepackage{lineno}
\usepackage{enumerate}
\usepackage{float}
\usepackage{subfig}
\usepackage{comment}

\begin{document}
\title{Dispersal-induced survival of predators in metacommunities due to transient chaos}

\author{Samali Ghosh}
\affiliation{Physics and Applied Mathematics Unit, Indian Statistical Institute, 203 B. T. Road, Kolkata 700108, India}

\author{Arnob Ray}
\affiliation{Indian Institute of Technology, Gandhinagar, Palaj, Gujarat 382055, India}

\author{Everton S. Medeiros}
\affiliation{Institute of Geosciences and Exact Sciences, São Paulo State University (UNESP), Avenida 24A 1515, 13506-900 Rio Claro, São Paulo, Brazil}

\author{Dibakar Ghosh}
\affiliation{Physics and Applied Mathematics Unit, Indian Statistical Institute, 203 B. T. Road, Kolkata 700108, India}

\author{Syamal Kumar Dana}
\affiliation{Centre for Mathematical Biology and Ecology, Department of Mathematics, Jadavpur University, Kolkata 700032, India}
\affiliation{Division of Dynamics,  Lodz University of Technology, Stefanowskiego 1/15, 90-924  Lodz, Poland}

\author{Tomasz Kapitaniak}
\affiliation{Division of Dynamics,  Lodz University of Technology, Stefanowskiego 1/15, 90-924  Lodz, Poland}

\author{Chittaranjan Hens}
\affiliation{Center for Computational Natural Science and Bioinformatics, International Institute of Informational Technology, Gachibowli, Hyderabad-500032, India}

\author{Ulrike Feudel}
\affiliation{Theoretical Physics/Complex Systems, Institute for Chemistry and Biology of the Marine Environment, Carl von Ossietzky University Oldenburg, 26129 Oldenburg, Germany}

\thanks{Corresponding author: ulrike.feudel@uni-oldenburg.de}

\begin{abstract}
Dispersal networks critically shape the fate of ecological communities, yet the
mechanisms linking connectivity and persistence remain poorly understood. We
show that an interplay between asymmetric dispersal and asynchronous dynamics across patches in a dispersal network can prevent predator extinction across broad dispersal ranges, even in identical environments in which synchrony usually drives ecosystems to collapse. Unlike classical rescue effects based on environmental heterogeneity or equilibrium states, this mechanism emerges from non-equilibrium dynamics, specifically from transient chaotic dynamics. Dispersal coupling perturbs local trajectories in patches facing extinction and reinforce chaotic motion, thereby sustaining chaotic oscillations indefinitely. Strikingly, only minimal connectivity is required: small-world networks with a few long-range links suffice to rescue predator populations. These findings reveal a counterintuitive principle that limited, well-placed connectivity can harness chaos to maintain biodiversity in fragmented landscapes.
\end{abstract}

\maketitle

\section{Introduction}
Species' extinctions have been a natural occurrence throughout Earth's history \cite{purvis2000extinction, jablonski2001lessons}, although recent extinction rates have accelerated due to human activity and other circumstances \cite{pimm1995future, ceballos2015accelerated}. Hence, one of the key challenges in contemporary ecology is preventing  further extinction of species by various conservation strategies
\cite{ricketts2005pinpointing, tilman2017future, bhatia2023network}. Key factors that play a large role in triggering extinction include the size of the population \cite{reed2003estimates,traill2007minimum}, climate change, and the effect of anthropogenic forcing like e.g. land use change \cite{kaiho2022relationship,duffy2022climate}. Extinction of species is manifested in the observed steady decline of the number of species in many different habitats over   

\maketitle


a very long time horizon \cite{lotze2006depletion} as well as in several mass extinction events which took place on Earth in the distant past \cite{sudakow2022knowledge}. 

\par Since the seminal work of May on the stability of complex ecosystems \cite{may1972will}, much theoretical and empirical research has been devoted to the question how species' interactions and connectivity between different landscapes shape diversity and stability in ecosystems \cite{leibold2004metacommunity, allesina2012stability, meena2023emergent}. While many studies have focused on equilibrium properties and feasibility conditions, ecosystems are often governed by non-equilibrium dynamics such as oscillations and chaos either manifested as stable attractors \cite{hastings1991chaos,  huisman1999competition, beninca2008chaos, blasius2020chaos} or as transient dynamics (chaotic saddles) \cite{morozov2020long, pattanayak2021bistability, morozov2024long}. These chaotic transients can dominate ecological time scales, making their understanding critical for predicting extinction risk \cite{hastings1994persistence}. 


\par Dispersal provides a potential buffer against such extinctions. It refers to the movement of organisms within or across communities in different habitats across the network of different patches of populations. By allowing organisms to move between patches, dispersal creates opportunities for recolonization and rescue effects and substantially influence the dynamics and stability of ecological systems \cite{holyoak1996persistence, vuilleumier2006does,bode2008using,vuilleumier2010effects,gilarranz2012spatial,shtilerman2015effects}. How dispersal can prevent extinction is related to the ability of the species to move to other patches to reach locations with better growth conditions. Additionally, dispersal offers the possibility of recolonization of habitats in which a species is close to extinction or is already extinct, constituting a rescue effect \cite{pulliam1988sources, loreau2003biodiversity}. While most of these works consider the stability of equilibrium solutions, the situation becomes more involved when each patch in the network possesses non-equilibrium dynamics like periodic or chaotic behavior. Then  synchronization of the population dynamics between different patches constitutes an additional complication, as it is considered to be detrimental, as extinction in one habitat would be leading to extinction in all habitats at once, and no rescue would be possible \cite{earn2000coherence, molofsky2005extinction}. For this reason, strong dispersal enhances the chances for synchronization and needs to be avoided. Therefore, to prevent extinction
of any species in any of its habitats, the establishment of moderate dispersal has to be realized, thereby averting complete synchrony between habitats \cite{holyoak1996persistence, holyoak2000habitat, holland2008strong, jansen2001dynamics,goldwyn2008can,hastings2010timescales,bjornstad1999spatial,goldwyn2009small}. Holland et al. \cite{holland2008strong} reported that an increment of randomness in a dispersal network (small world regime or beyond) is more favorable than a regular network of non-local structure for prolonged transient dynamics. Additionally, their findings emphasize that desynchronization or cluster synchronization favors species persistence. Medeiros et al. \cite{medeiros2021impact} showed that heterogeneous dispersal, i.e., different dispersal strength, can suppress chaotic dynamics leading to the stabilization of equilibrium solutions.

\par Here, we study the interplay between dispersal and transient non-equilibrium dynamics to show that dispersal helps maintaining biodiversity in a metacommunity by stabilizing transient chaotic dynamics to prevent extinction of a top predator. We show that 
asynchronous dynamics between the different patches are a necessary condition for avoiding extinction. The transient chaotic dynamics is turned into a long transient with extraordinary long life time or even into a chaotic attractor since dispersal reinforces chaos by keeping trajectories within the chaotic set even in parameter ranges in which the chaos would disappear, and with it the top predator. We demonstrate the mechanism how dispersal can prolong the lifetime of such chaotic transients, transform them into sustained chaotic attractors, and thereby rescue predator populations otherwise destined for collapse, and how these rescue effects  depend on the strength of dispersal, the topology of the network, and the underlying nonlinear interactions that drive the dynamics. Addressing this gap is crucial, as many real ecosystems operate not near equilibria but within transient-dominated regimes, where biodiversity may hinge less on the stability of equilibria than on the capacity of long-lived chaos to delay or even avert extinction. We illustrate this mechanism of dispersal-induced survival by considering the dynamics of a three-dimensional resource-consumer-predator model originally introduced by Hastings et al. \cite{hastings1991chaos} and further studied in different contexts \cite{pattanayak2021bistability,meng2025dispersal}. 


We consider twenty coupled  patches each of them following the dynamics of the resource-consumer-predator model in a metacommunity modelled as three disparate dispersal networks  as globally coupled, random, and small-world topologies. By increasing the dispersal rate or the coupling strength, we observe that the predator population of all patches leads to prolonged transient dynamics (or a permanent chaotic state), irrespective of the topology of the dispersal networks. Using a global stability measure based on the concept of basin stability, our study validates that a non-local topology, augmented with a small proportion of extra links, proves more advantageous for the persistence of the predator population, compared to patches connected globally or randomly but with more connectivity. We clarify the mechanism of how these long transients emerge and demonstrate that only  synchronization of the patches finally leads to extinction of the predator. 

The remaining parts of this article are organized as follows: In Sec.~\ref{section2}, we present the mathematical framework for a prey-predator metacommunity, along with the construction of the networks. We show in Sec.~\ref{section3} how this chaotic dynamics containing all three species is stabilized due to dispersal and leads to very long transients or even to a chaotic attractor for the three different network types (fully connected, random, and small world). Section~\ref{section4} is devoted to the stability analysis and the robustness of the phenomenon. Finally, the mechanism of how this extreme prolongation of the chaotic transients is achieved is discussed in Sec.~\ref{section5}. We discuss our results finally in Sec.~\ref{section6} and compare them with other approaches in the literature. 

\section{Model description and network construction}\label{section2}
\subsection{Model description}\label{model}

During the 1990s, theoretical ecologists developed numerous two or three-species models to comprehend the phenomena of extinction and survival of species within natural ecosystems \cite{rosenzweig1973exploitation,hastings1991chaos}. In our study, we consider a three-trophic food chain model consisting of resource ($R$), consumer ($C$), and predator ($P$) interactions which is based on a previous predator-prey model proposed by Hastings and Powell \cite{hastings1991chaos} with a parameterization of Yodzis \& Inne  \cite{yodzis1992body} and McCann \& Yodzis \cite{mccann1994nonlinear}, respectively. In the absence of any consumers, the model assumes that the resource grows logistically until it reaches the carrying capacity. Both the consumer and the predator evolve according to a Holling type II functional response, i.e., they both die out in the absence of their respective resources; their consumption rate reaches saturation for large resource or prey concentrations, respectively. In our model, we look at the dynamics of a metacommunity made up of $N=20$ patches with diffusive interaction restricted to consumers and predators, mimicking their dispersal between the patches. The dynamics of the species, including the dispersal of consumers and predators, is described by a set of differential equations in dimensionless form:

\begin{align}\label{eq1}
\dot{R}_i =& R_i \left(1 - \dfrac{R_i}{K}\right) - \dfrac{x_c y_c C_i R_i}{R_i + R_0},\nonumber \\
\dot{C}_i =& x_c C_i \left(\dfrac{y_c R_i}{R_i + R_0} - 1\right)-\dfrac{x_p y_p C_i P_i}{C_i + C_0} \nonumber \\
&+\dfrac{\epsilon_c}{d_i}\sum_{j=1}^{N} A_{ij}\left(C_j - C_i\right), \nonumber \\
\dot{P}_i =& x_p P_i \left(\dfrac{y_p C_i}{C_i + C_0} - 1\right) + \dfrac{\epsilon_p}{d_i} \sum_{j=1}^{N} A_{ij}\left(P_j - P_i\right),
\end{align}
where $i=1,2,\cdots, N$ denotes the patch index and $R_{i}, C_{i}, P_{i}$ denote the resource, consumer, and predator population density in the $i$-th patch, respectively. The parameters $x_i, i = {c,p}$ are the mass-specific metabolic rates of the consumer and predator, respectively, while $y_i, i = {c,p}$ are measures of the ingestion rates per unit metabolic rate of the species $C$ and $P$. $R_0$ and $C_0$ are the half-saturation constants. The only parameter which is accessible for a variation is the carrying capacity $K$, which is chosen to be identical in each patch. The consumer depends on the resource for its existence, and the predator depends on the consumer. Consumers and predators move from one patch to another with dispersal strengths $\epsilon_c$ and $\epsilon_p$, respectively. All patches and their interconnections by dispersal are arranged in a network in which the nodes correspond to the patches and the edges are the dispersal paths. The adjacency matrix of the dispersal network of the metacommunity is represented by $A_{ij}$, which determines the connections between the $i$-th and $j$-th patch. Specifically, if dispersal takes place between the $i$-th and $j$-th patches, then $A_{ij}=1$; otherwise, $A_{ij}=0$. The degree of the $i$-th patch $d_{i}$ represents the number of patches connected with the $i$-th patch. Here, the dispersal rates are $\dfrac{\epsilon_c}{d_i}$ and $\dfrac{\epsilon_p}{d_i}$ for the $i$-th patch, respectively. According to \cite{yodzis1992body}, we set the parameters $x_c=0.4,\ y_c=2.009$ for the consumer, and $x_p=0.08,\ y_p=2.876$ for the predator, ensuring that both are either invertebrate or vertebrate ectotherms, thereby maintaining a realistic predator-prey body mass ratio. According to \cite{mccann1994nonlinear}, we fix $R_0=0.16129$, $C_0=0.5$. We numerically integrate our system for $T=10^5$ (arb. units) with step-size $dt=0.01$ using a $4$-th order Runge-Kutta method.

\begin{figure*}[!ht]
	\centering
	\centerline{\includegraphics[scale=0.7]{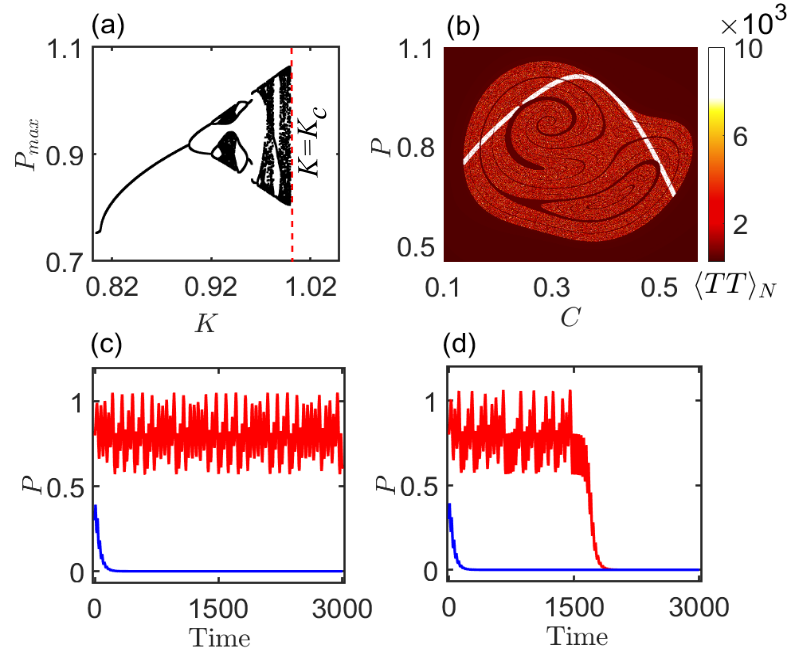}}
    \caption{Dynamical behavior of the single $R$--$C$--$P$ model. (a) Bifurcation diagram of $P$ as a function of $K$, shown in terms of local maxima of $P$. The system's chaotic behavior emerges through a period-doubling cascade, and the chaotic attractor disappears after crossing the critical parameter value $K_c$. The co-existing limit cycle in the $R$--$C$ plane is not plotted here since $P=0$. (b) Transient time distribution for the single $(R, C, P)$ model at $K=1.0$. The white dots represent the chaotic saddle in terms of a Poincaré section at $R = 0.5$. (c) Temporal evolution of the predator population for the pre-crisis ($K=0.99$) and (d) for the post-crisis ($K=1.0$) regime, respectively. The red and blue trajectories differ only by the choice of the initial conditions. Initial conditions for blue trajectory: $(R, C, P)=(0.5, 0.4, 0.3)$ and for the red trajectory $(0.5, 0.3, 0.8)$.} 
    \label{fig1}
\end{figure*}

We first investigate the dynamics of the uncoupled food chain, i.e., $\epsilon_c = \epsilon_p =0$. Since the carrying capacity $K$ specifies the conditions of the environment, it serves as a meaningful bifurcation parameter. In Fig.~\ref{fig1}(a), we represent the bifurcation diagram where the maximum population density of the predators, $P_{\rm max}$, is shown as a function of $K$. We observe that the stable regular dynamics of the food chain, containing all three species, undergo a period-doubling route to chaos. In addition to this scenario, the state space of the isolated food chain possesses an alternative stable solution: a limit cycle on the $R$-$C$ plane with only populations of resource and consumer species; the predator species is extinct. Interestingly, at a critical value of $K$ ($K=K_c\approx 0.99976)$, the chaotic attractor collides with the boundary of its own basin of attraction in a boundary crisis (see dashed line in Fig.~\ref{fig1}(a)) \cite{grebogi1982chaotic, grebogi1983crises,ray2020another}. In this process, the chaotic attractor loses its stability and is converted into a chaotic saddle \cite{lai2011transient}, i.e. a chaotic dynamics which is unstable such that trajectories on this chaotic saddle will eventually escape from it reaching another attractor. We estimate the transient time for trajectories to converge to the alternative attractor -- the limit cycle on the $R$--$C$ plane -- as follows \cite{kittel2017timing, ray2021mitigating}:
\begin{eqnarray}
  T = inf\{t: |P(t)| < \delta\},
  \label{Eq:transient}
\end{eqnarray}
where $\delta=0.001$ is a prescribed threshold to numerically specify the extinction of the predator species, indicating the trajectory's approach to the stable limit cycle. In Fig.~\ref{fig1}(b), the color code shows $T$ obtained for a grid of initial conditions on the $C(0)$--$P(0)$ plane with $R$ 
fixed at $R(0)=0.5$. The uniform dark red regions correspond to trajectories that converge to the limit cycle without approaching the chaotic saddle. By contrast, the initial conditions color-coded from light red via yellow to white form a fractal structure of trajectories that approach the chaotic saddle, oscillate chaotically, and then escape towards the limit cycle. In this figure, the chaotic saddle denoted by $\Lambda$ is shown as white dots. The probability distribution of durations of the chaotic transients observed in Fig.~\ref{fig1}(b) follows an exponential distribution \cite{lai2011transient}. In addition, the fractal structure on display in this figure 
corresponds to an approximation of the stable foliation, loosely speaking, the stable manifold of the chaotic saddle $\Lambda$.
To demonstrate such bistability, we first fix the carrying capacity at $K=0.99 <K_c$ in Fig.~\ref{fig1}(c) and show the time evolution of the density of the predator species for two sets of initial conditions: trajectories starting in the basin of attraction of the chaotic attractor are depicted in red, and trajectories starting in the basin of attraction of the limit cycle are depicted in blue. Indeed, the food chain dynamics are three-dimensional, involving the predator species in the chaotic attractor, while the dynamics in the limit cycle are restricted to the $R$--$C$ plane. In the boundary crisis, the basin of attraction of the former chaotic attractor disappears. As a result, the initial conditions that converged to the chaotic attractor in the pre-crisis scenario ($K<K_c$) now exhibit transient chaotic dynamics in the post-crisis scenario ($K>K_c$) (Fig.~\ref{fig1}(d)), within a finite time interval during which trajectories are close to the chaotic saddle. Therefore, the predator species is present only for this finite time interval corresponding to the chaotic transients. 
Subsequently, the trajectory escapes from the vicinity of the chaotic saddle and converges to the limit cycle on the $R$-$C$ plane as the sole stable solution in the post-crisis regime, with the predator population going eventually extinct \cite{hastings2018transient}.

According to our aim to study the rescue of the predator due to dispersal, we now specify the connectivity properties of our networks, detailing the conditions in which the interplay between the dispersal among patches and the local dynamics of the food chain in the post-crisis ($K>K_c$) regime counterbalances the boundary crisis, thereby improving species survivability in the metacommunity model.

\subsection{Construction of the network architecture}
\begin{figure*}
\centering
	\centerline{\includegraphics[scale=0.5]{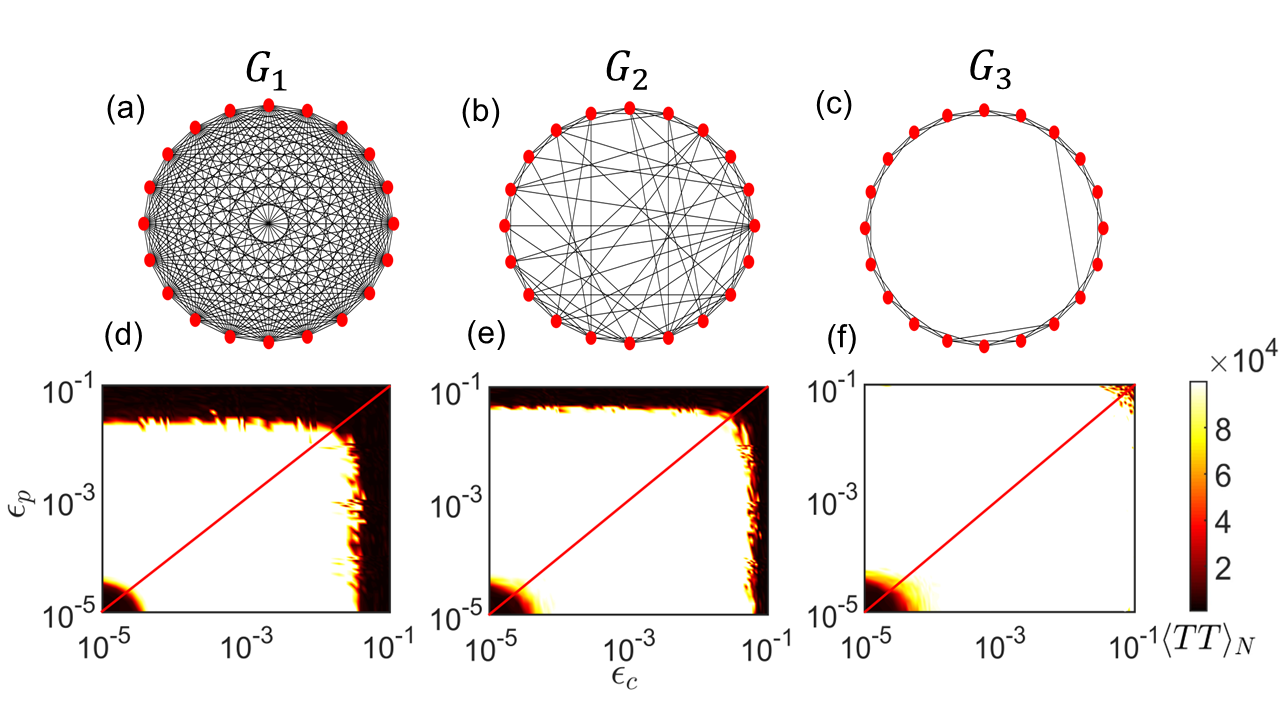}}
\caption{(a)-(c) Three representative network topologies with $N=20$ patches, connected according to structures $G_1$, $G_2$ and $G_3$.
(d)-(f) Average transient times $\langle TT \rangle_{N}$ before reaching the attractor depending on dispersal strengths in the  $\epsilon_c-\epsilon_p$ plane within the range $[10^{-5},10^{-1}]$. The white region indicates a stable, chaotically oscillating predator population over a long time ($\sim10^{5}$ time units), without extinction; The lower black region denotes the short-lived transient period preceding the predator's extinction while the upper black region in (d) and (e) indicates the sharp coherent extinction of the predator population in all patches. Notably, the white region is significantly enhanced for $G_2$ and $G_3$ when the number of links is substantially decreased compared to the network ($G_1$). The $\epsilon_c$=$\epsilon_p$ line is shown in red.}
\label{fig2} 
\end{figure*}

To study the role of dispersal in the metacommunity model, we consider three network topologies. Here, we investigate a non-local network with $N=20$ patches, where each patch is connected with its four nearest neighbors. We add links or {\it shortcuts} randomly between any two patches with a certain probability, say $p$. Using this method, we construct two different network topologies of a random network by setting $p=1$ and $p=0.01$, respectively. This construction is based on the {\it Newman and Watts} algorithm \cite{newman1999renormalization,newman2003structure}. In our study, we deal with the following three types of network topology in where we denote as topology I a globally connected network $G_1$, as topology II a random network $G_2$ with $p=1$, and as topology III another random network, called a small world network, $G_3$ with a much smaller probability $p=0.01$, creating additional long-range connections. The latter corresponds to a small-world network in which the majority of connections are local with a certain radius of locality, with a few long-range connections added. The fully connected network $G_1$ consists of $1225$ links, while the topology of the two random networks are much sparser. The link density of the network $G_2$ is almost two times larger (no. of links is $80$) than that of the network $G_3$ (no. of links is  $42$). The average shortest path lengths of $G_1$, $G_2$, and $G_3$ are $1.0$, $1.589$ and $2.621$ respectively and the clustering coefficients of the networks are $1.0$, $0.452$ and $0.49$, respectively. Figs.~\ref{fig2}(a)-(c) represent these networks. Our motivation for this choice of network topologies focuses on how the number of connections in a metacommunity influences the extinction of the predator population for each patch.

\begin{figure*}[ht]
\centering
	\centerline{\includegraphics[scale=0.5]{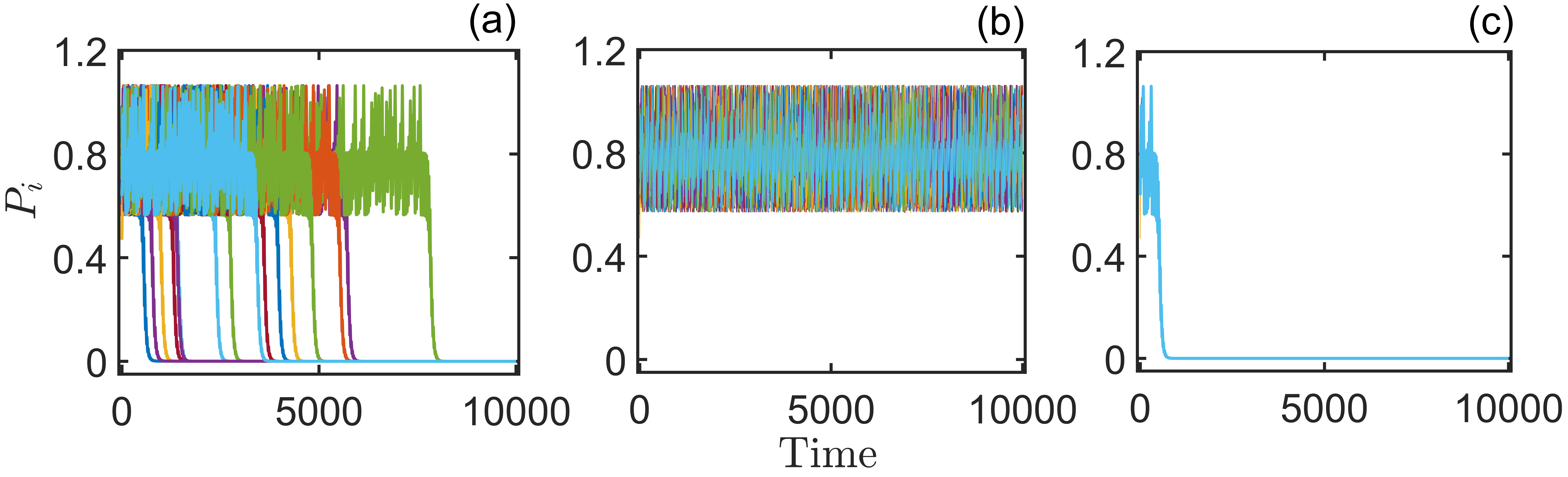}}
\caption{Time signals of predator populations in $G_1$ for three different coupling strengths. (a) For $\epsilon_p=2\times 10^{-5}$, we plot $P_i, i=1,2,...,20$, exhibiting transient chaos, and each patch, plotted in a different color, shows a different transient time. (b) For $\epsilon_p=10^{-3}$, a long transient chaotic phase for each patch is observed, leading to permanent chaos (it is checked that the persistence of chaotic oscillation lasts at least $10^{5}$ time units for all oscillators). Please note that they oscillate in a desynchronized pattern. (c) All patches show a short transient chaotic phase, and interestingly, all predators are extinct in a synchronized manner for $\epsilon_p=10^{-1}$. We fix the parameter $\epsilon_c=  10^{-5}$ in all figures.} 
    \label{fig3} 
\end{figure*}

\section{Dispersal-induced survival of the predator species}
\label{section3}

We now demonstrate the phenomenon by which the predator species, which occurs only for a transient time in the uncoupled food chains, can survive indefinitely in the diffusively coupled dispersal network. At first, we prescribe the initial conditions of every patch in our networks such that they approach the chaotic saddle $\Lambda$ in their isolated state space before escaping to the limit cycle on the $R$-$C$ plane, which has no predators. With this setup, the predators go extinct at different time instants for each one of the different patches. To quantify those transient times for the coupled case, we extend the definition of {\it transient times} in Eq.~(\ref{Eq:transient}) to account for the predator population in the $k$-th patch as $T_k = inf\{t: |P_{k}(t)| < \delta\}$, where the threshold is set again to $\delta=0.001$. Subsequently, to quantify the duration of transients in the entire network, we define the {\it mean transient time} over all network patches as $\langle TT \rangle_{N}=\dfrac{\sum_{k=1}^{N} T_k}{N}$. Hence, in the $\epsilon_c$-$\epsilon_p$ parameter plane shown in Figs.~\ref{fig2}(d)-(f), we depict the numerical calculation of $\langle TT \rangle_N$ for the networks $G_1$, $G_2$, and $G_3$ by varying both $\epsilon_c$ and $\epsilon_p$ in the range $[10^{-5},10^{-1}]$. In these diagrams, the color code indicates the value of $\langle TT \rangle_N$ for the corresponding parameter pair. In addition, we draw the $\epsilon_c=\epsilon_p$ line in red to distinguish the two regions: $\epsilon_p > \epsilon_c$ and $\epsilon_p < \epsilon_c$.

Since in nature the predator dispersal rate is generally higher than the one of the consumer \cite{otto2008predator}, we focus our attention on the region $\epsilon_p > \epsilon_c$ (the upper left triangle of the diagonal in Figs.~\ref{fig2}(d)-(f)). Hence, if the predator persists in {\it each} patch throughout the entire numerical simulation without reaching zero, we consider the entire system as stable, meaning it can avoid extinction. We call the system persistent if the transient time is longer than the entire simulation time $\langle TT \rangle_N=10^5$ and indicate it with white color. The black color indicates transient times leading to predator extinction. For small dispersal strengths $\epsilon_c$ and $\epsilon_p$ (in the lower black regime), the predator population in each patch $k$ has a different transient time $T_k$  that eventually leads to its extinction. This dynamics is visualized in Fig.~\ref{fig3}(a), where the evolution of the predator population in the $20$ patches is displayed. Even though the transient dynamics is chaotic in this regime, the predators go extinct and trajectories end up in different phases of the limit cycle on the $R$--$C$ plane. Conversely, in the white region of the diagrams in Figs.~\ref{fig2}(d)-(f)), the predator population of every patch avoids extinction (confirmed by the time evolution in Fig.~\ref{fig3}(b)), i.e., each of the predator populations persists throughout the entire course of numerical simulation. Please note, that the different colors of the dynamics in different patches indicate that the patches oscillate in asynchrony. Another black region is marked close to the top
of the  $\epsilon_c-\epsilon_p$ parameter plane (shown in Figs.~\ref{fig2}(d)-(e)), leading to the extinction of predator populations. The corresponding time evolution of this regime of large dispersal strengths is shown in Fig.~\ref{fig3}(c). It shows that the population of predators in all patches goes extinct in a completely synchronized manner, all at the same time.

\begin{figure*}
   \centering
	\centerline{\includegraphics[scale=0.5]{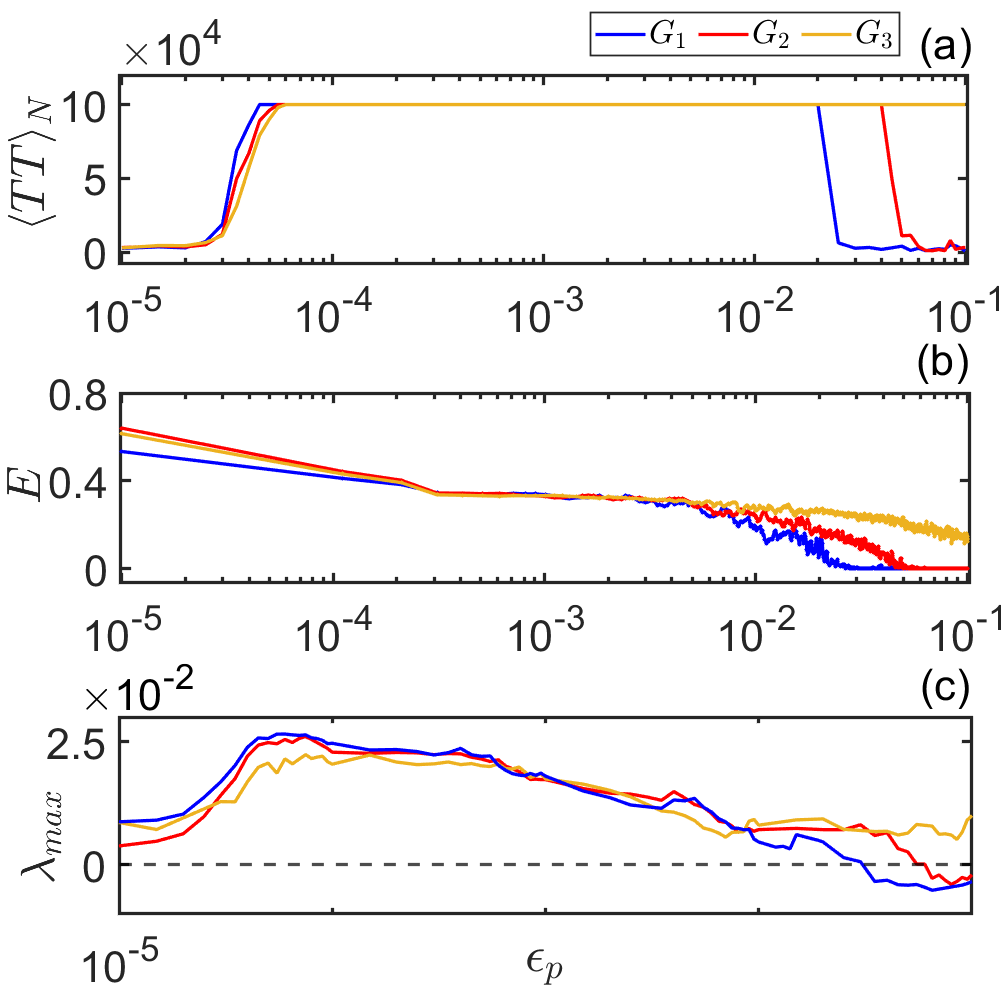}}
    \caption{(a) Mean transient time ($\langle TT \rangle_N$) versus $\epsilon_p$ in the presence of diffusive dispersal among consumers and predator populations. We plot $\langle TT \rangle_N$ by varying $\epsilon_p$ for $G_1$ (blue), $G_2$ (red) and $G_3$ (yellow). When $\langle TT \rangle_N =10^5$, the extinction of the predator population is prevented in every patch. (b) Time average synchronization error $E$ versus $\epsilon_p$. $E=0$ demonstrates the synchronization of the metacommunity dynamics, including all $20$ patches. (c) The maximum Lyapunov exponent varies with $\epsilon_p$. The maximum Lyapunov exponent of the system is positive when the system evolves in desynchrony and becomes negative when the whole metacommunity turns to synchronization for all three types of network coupling topologies. We fix $\epsilon_c$ at $10^{-5}$ for the entire figure.} 
\label{fig4} 
\end{figure*}

Our investigation suggests that, besides the dispersal strength, the network structure also plays an important role in the survival of predator populations. Noticeably, the white region of the parameter space is significantly enhanced for the network $G_{3}$ as shown in Fig.~\ref{fig2}(f). This network (topology III, $G_3$), consisting of $42$ links (where only two random links are added in the existing ring), shows a wider region of white space than $G_2$, consisting of $80$ links, and $G_1$, consisting of $1225$ links. As a result, we conclude that an asymmetric network, adding a tiny fraction of links to an existing ring, produces the most beneficial outcomes in which biodiversity is maintained for a wide range of dispersal rates of predators. By contrast, our results for topologies $G_1$ and $G_2$ suggest that a large ``mixing" among patches in the metacommunity in the sense of large dispersal due to a larger number of connections, may disturb the stability of the whole system as it leads to synchronization of all patches, making them prone to the extinction of the predator. Another intriguing finding is the strong difference between the collective behavior of predators in the lower and the upper black regimes. While the top black regime for large $\epsilon_p$ shows that the predators die out synchronously, the weaker dispersal strength of the predator (lower black region) reveals a desynchronized chaotic transient. For intermediate dispersal strength, the chaotic oscillations are desynchronized (white space). These observations support one of ecology's central hypotheses, {\it synchrony does not favor diversity} \cite{zhang2018critical}, and offer a surprising finding: networks possessing only a few dispersal corridors may avoid extinction more efficiently than well-connected patches of populations.
  
Next, we fix $\epsilon_c$ at $=10^{-5}$ and vary $\epsilon_p$ in the interval $[10^{-5},10^{-1}]$ to compare the efficiency of the network topologies leading to the survival of the predator population. At first, we show the mean transient times ($\langle TT \rangle_N$) for three types of network topologies (blue for $G_1$, red for $G_2$, and yellow for $G_3$) in Fig.~\ref{fig4}(a). Beyond a certain critical value of $\epsilon_p$, $\langle TT \rangle_N$ is raised almost four orders of magnitude, and it reaches the value of $10^5$. It indicates that all patches of the metacommunity show the survival of the predator population. The critical onset of such survivability remains almost the same for all network topologies. However, the critical value at which survivability ceases varies significantly for the different topologies. Particularly, the extinction occurs faster ($\langle TT \rangle_N$ drops early) for $G_1$ compared to $G_2$ and $G_3$ with respect to $\epsilon_p$. Therefore, we can conclude that the interval of dispersal strength $\epsilon_p$ within which survival of predators occurs widens as the number of links decreases. We denote the length of the interval of $\epsilon_p$ as $\Delta \epsilon_p$ for which all predator populations of the metacommunity survive up to $10^{5}$ time units. We numerically calculate $\Delta \epsilon_p$ which is the difference between two end points of $\epsilon_p$ for which $T_1=T_2=...=T_{20}=10^5$. We find that $\Delta \epsilon_p \approx 0.029975$ for $G_1$, $\Delta \epsilon_p \approx 0.059975$ for ${G_2}$ and, $\Delta \epsilon_p \approx 0.099975$ for $G_3$, i.e., $\Delta \epsilon_p|_{G_3}> \Delta \epsilon_p|_{G_2} > \Delta \epsilon_p|_{G_1}$, within the considered interval of $\epsilon_p$ is $[10^{-5},10^{-1}]$ in this numerical experiment.

\section{Stability analysis of the completely synchronized state}
\label{section4}

From our previous results, it is noticeable that different types of dynamics occur when the dispersal rate of the predator $\epsilon_p$ is varied. While for a low dispersal rate the predator species goes extinct and the dynamics becomes periodic with frequency synchronization among the patches (Fig.~\ref{fig3}(a)), the predator persists in asynchronous chaotic dynamics of the patches for a moderate dispersal rate, as shown in Fig.~\ref{fig3}(b). Finally, for a large dispersal rate, complete synchronization in the limit cycle on the $R$--$C$ plane occurs (Fig.~\ref{fig3}(c)). In this context, we aim to gain further insights into the stability of our system. To this end, we examine the long-term behavior of the system dynamics in terms of the synchronization patterns across all patches. We start by defining the synchronization error for the patches in our system as:
\begin{equation}\label{eq2}
    E=  \left \langle \sum_{i=1}^{N} \dfrac{\sqrt{(R_i-R_1)^2+(C_i-C_1)^2+(P_i-P_1)^2}}{N(N-1)} \right \rangle_t, 
\end{equation}  
where $\langle \cdots \rangle_t$ stands for the time average over a long time interval $t$ after disregarding initial transients. Following this definition, we set $\epsilon_c=10^{-5}$ and calculate the synchronization error by varying $\epsilon_p$ in the interval $\epsilon_p \in [10^{-5},10^{-1}]$ for the three different network topologies, as illustrated in Fig.~\ref{fig4}(b). We observe that the synchronization error $E$ for $G_1$ (depicted by a blue curve) goes to almost zero for smaller values of $\epsilon_p$ than for $G_2$ (depicted by the red curve). Here, a critical transition occurs between an asynchronous chaotic state and a synchronous limit cycle state in the respective metacommunity. The synchronization error $E$ for $G_3$ (depicted by the yellow curve) never reaches zero in the range $\epsilon_p \in [10^{-5}, 10^{-1}]$. Near the critical value of $\epsilon_p$, the completely synchronized state may or may not occur. In order to confirm this possibility, we perform a linear stability analysis of the completely synchronized state using the master stability function approach \cite{pecora1998master}, which we discuss in the following subsection.

\subsubsection{Linear stability analysis of the synchronized state: Master stability function approach}

In general, in a system of coupled oscillators, the interaction strength among the oscillators plays an important role in determining their collective dynamics. As the coupling strength decreases, the ability of the oscillators to synchronize their behavior diminishes. Weakening synchrony in the weak coupling regime reduces the level of influence and coordination among the oscillators. In our system, the interaction strength corresponds to the strength of dispersal between the patches, while the synchrony refers to the synchrony in dynamics across patches. The Master Stability Function (MSF) is a concept used to analyze the stability of networked dynamical systems, particularly in the context of synchronization. By applying the master stability framework, we can assess the stability of a networked system across different parameter values, identify critical points where stability transitions occur, and predict the emergence of synchronization patterns in complex networks. To determine a suitable coupling range for our model \eqref{eq1}, we take the help of the MSF approach \cite{arenas2008synchronization, huang2009generic} and find the values of $\epsilon_p$ for which complete synchronization occurs in the metacommunity. 

To study the stability of the synchronous state, it is sufficient to observe the maximum Lyapunov exponent $\lambda_{max}$, which we can calculate numerically (see \cite{arenas2008synchronization,huang2009generic} for details). If $\lambda_{max}$ is less than zero for a dispersal network, all patches become synchronized.

In Fig.~\ref{fig4}(c), we depict the variation of $\lambda_{max}$ by varying $\epsilon_p$ in the interval $\epsilon_p \in [10^{-5},10^{-1}]$. We identify two critical values of $\epsilon_p$, beyond which complete synchronization becomes stable in the dispersal networks $G_1$ and $G_2$. It is important to note that the completely synchronized state corresponds to the stable solution of an isolated food chain. Consequently, this state follows a limit cycle within the $R$-$C$ plane, characterized by the absence of the predator species. Building on this understanding of the linear stability of the synchronized state, we complement this analysis by gathering information about the global stability of the completely synchronized state. This analysis is useful in determining whether the completely synchronized state coexists with other desynchronized solutions within the system's state space. To conduct this analysis, in the next subsection, we use the concept of Basin Stability (BS), which estimates the relative volume of a basin of attraction \cite{menck2013basin}.

\subsubsection{Global stability analysis of the synchronized state: Basin stability approach}
\label{Basin stability} 

Lyapunov exponents offer insights into the local stability properties around trajectories or attractors. Specifically, a negative Lyapunov exponent implies local stability, indicating that small perturbations diminish over time and nearby trajectories converge locally to the attractor. By contrast, Basin Stability (BS) assesses global stability by measuring the relative volume of the state space from which initial conditions converge to a particular attractor or synchronization state. Hence, BS provides a broader perspective on the robustness of dynamical states against substantial perturbations. In our study, the considered model exhibits multi-stability, which aligns with previous work by Pattanayak et al. \cite{pattanayak2021bistability}, who illustrated how the basin of attraction varies with changing bifurcation parameters. Motivated by this finding, we investigate the effect of the choice of initial condition for different network topologies on the stability that can arise due to interactions among models. Basin stability analysis explicitly evaluates robustness against large perturbations or diverse initial conditions, making it particularly relevant for ecological systems commonly subjected to substantial disturbances and inherent variability. Although a negative Lyapunov exponent indicates local stability of synchronized states, it does not necessarily imply global stability; multiple attractors, including both synchronous and asynchronous states, may coexist. Therefore, basin stability analysis is crucial, as it quantifies the global significance of these attractors by evaluating the likelihood that trajectories initiated from random initial states will converge to synchronization. Consequently, BS effectively differentiates local stability from global one.

Suppose the system in Eq.~\eqref{eq1} converges to the synchronous state for a number $I_S$ of initial conditions. We estimate the BS for the synchronous state as $\dfrac{I_S}{I}$, where $I$ is the total number of initial conditions chosen from the state space in a numerical experiment. The value of BS ranges from $0$ to $1$. BS = $0$ implies the synchronized state remains unstable for all initial conditions, while BS = $1$ indicates that the synchronized state is globally stable. If $0 <$ BS $< 1$, the probability of arriving at the synchronous state for some initial conditions is non-zero. As an example, employing BS analysis, we aim to demonstrate the presence of only a limited range of dispersal strengths where a non-zero ($0 <$ BS $< 1$) fraction of initial conditions leads to complete synchronization.

\begin{figure*}[ht]
\includegraphics[width=\linewidth]{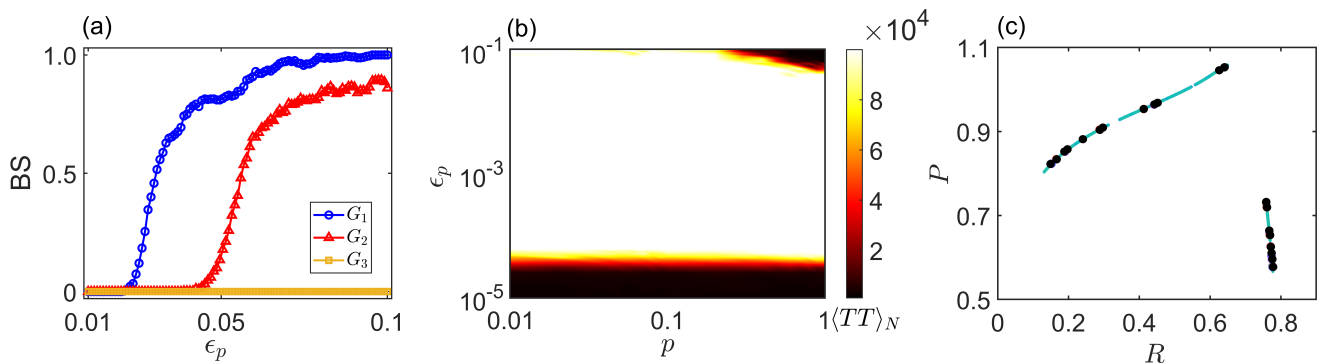}
\caption{(a) Basin stability (BS) measurement. We measure BS for the networks $G_1$ (blue), $G_2$ (red), and $G_3$ (yellow) with respect to the dispersal rate of the predator $\epsilon_p$. We identify appropriate initial conditions to achieve a synchronous state for $G_1$ and $G_2$, except for $G_3$. $\epsilon_c$ is fixed at $10^{-5}$. We calculate BS for each point by taking the mean over $1000$ realizations. (b) $\epsilon_p$--$p$ parameter space for showing the variation $\langle TT \rangle _N$. We depict the variation of mean transient time for different dispersal networks for a range of $\epsilon_p \in [10^{-5}, 10^{-1}]$. We vary $p \in [0.01, 1]$ (probability for generating the network topology), and for each value of $p$, we construct $10$ networks and choose the realization of $10$ initial conditions for each simulation associated with a network. $\epsilon_c$ is fixed at $10^{-5}$. (c) Snapshot of the network trajectories at $t = 10^5$ (arb. units). The circles marked in black represent patches of the coupled system. The chaotic saddle $\Lambda$ in the state space of the patches is shown by the cyan dots, corresponding to a Poincaré section with the consumer biomass fixed at $C = 0.4$. The network contains $N = 20$ patches arranged in the $G_3$ topology, with coupling strengths $\epsilon_p = 10^{-3}$ and $\epsilon_c = 10^{-5}$. The carrying capacity is $K = 1.0$.} 
\label{fig5} 
\end{figure*}

Specifically, to measure the BS, we numerically integrate the system in Eq.~\eqref{eq1} for $\textit{I} = 1000$ initial conditions selected from a region of the state space where the predator populations persist for times longer than $10^5$ (arb. units), as observed in Fig.~\ref{fig4}. Hence, Fig.~\ref{fig5} shows the variation of BS by changing the dispersal rate of the predator $\epsilon_p \in [0.01,0.1]$ where the transition from desynchronization to a synchronous state is possible to occur as shown in Fig.~\ref{fig4}. For the lower strength of dispersal, e.g., $\epsilon_p = 0.01$, we observe that $BS=0$ for $G_1, G_2$ and $G_3$, indicating that all initial conditions do not approach the synchronized state. In fact, for this dispersal strength, all the patches exhibit desynchronous chaotic dynamics, with predator species persisting for all three topologies. However, further increasing $\epsilon_p$ reveals different behaviors across different network topologies. For instance, in topology $G_1$, BS (blue curve) becomes non-zero at around $\epsilon_p=0.02$, which is smaller compared to the other topologies. This indicates that, at this dispersal rate, a portion of the initial conditions tends to achieve complete synchronization, with no predator species present for $G_1$. In addition, at a dispersal rate around $\epsilon_p = 0.1$, the BS reaches the value of 1, indicating that the synchronized state becomes globally stable. For topology $G_2$, BS (red curve) becomes non-zero at a slightly higher dispersal rate ($\epsilon_p \approx 0.045$) compared to $G_1$. Similar to $G_1$, BS increases significantly with increasing $\epsilon_p$. Finally, for topology $G_3$, we observe that BS (yellow curve in Fig.~\ref{fig5}(a)) remains zero over the entire considered interval of $\epsilon_p$. Therefore, for this topology, all initial conditions lead to desynchronous chaotic dynamics with survival of the predator species. This result is in agreement with the linear stability analysis of the synchronized state shown in Fig.~\ref{fig4}(c).

\subsubsection{Robustness of predator survivability to changes in network topology}
\label{Robustness of prolonged transient}

To complete our analysis, we investigate the robustness of the survivability of predator species against random selection of the network architecture. All the results above are depicted for three particular networks ($G_1$, $G_2$, and $G_3$).

To check the robustness of our results, we measure the variation of the mean transient time $\langle TT \rangle_N$ of the predator survival by simultaneously varying the probability $p$ of randomly adding a network link and the coupling strength $\epsilon_p$. Specifically, we vary $\epsilon_p$ within $[10^{-5}, 10^{-1}]$ and generate $10$ distinct networks, each constructed using the Newman and Watts algorithm with various probability values $p \in [0.01, 1]$. For each value of $p$, we consider $10$ distinct networks and $10$ sets of initial conditions. This analysis is depicted in Fig.~\ref{fig5}(b) with the same color coding as Figs.~\ref{fig2}(d)-(f). We verify that the phenomenon of predator survival is robust over a large interval of the probability $p$ (white color in Fig.~\ref{fig5}(b)). However, for large values of $p$, corresponding to a more densely connected network, the interval of $\epsilon_p$ for predator survival shortens. This observation is consistent with the results for $p=0.01$ and $p=1$ shown in Figs.~\ref{fig2}(e)-(f).

\section{The mechanism for the dispersal-induced survival of the predators}\label{section5}

We now shed light on the dynamical mechanism that ensures the survivability of predator species in the metacommunity. To this end, we first recall that, in the description of the isolated food chain in Sec.~\ref {section2}, the predator species only persists for a finite time corresponding to the time interval in which a trajectory is in the vicinity of the chaotic saddle $\Lambda$. In general, the time intervals observed for an ensemble of trajectories to escape the vicinity of such chaotic saddles are exponentially distributed \cite{lai2011transient}. For diffusively coupled systems, such as our dispersal-coupled patches, the average escape time of trajectories leaving the chaotic saddle is expected to increase with the number of units that compose the dispersal-coupled system \cite{Crutchfield1988, lai2011transient, Wolfrum2011}. Nevertheless, even in the dispersal network, the escape times of an ensemble of trajectories still follow probability distributions due to different maximum proximity to the chaotic saddle. This behavior is indeed observed in the time evolution of the different patches for the dispersal rate $\epsilon_p=2\times 10^{-5}$, as shown in Fig.~\ref{fig3}(a). However, as depicted in Fig.~\ref{fig3}(b), the predator species persists equally for long times for many trajectories, not exhibiting a scaling for the escape times. This observation suggests that the mechanism underlying the predator's survival discussed here is different from the typical growth of chaotic transients with the system size, i.e., the number of dispersal-coupled population patches. Instead, the system trajectories remain indefinitely in the vicinity of the chaotic saddle. To illustrate the trapping of trajectories in the vicinity of the chaotic saddle, transforming it seemingly to an attractor or an extraordinary long chaotic transient, we show in Fig.~\ref{fig5}(c) a snapshot at $t=10^5$ (arb. units) of the resource and predator population densities of all patches (black full circles) in our system, overlaid onto the chaotic saddle $\Lambda$ (cyan dots).

This trapping mechanism in the chaotic saddle arises from the interplay between the dispersal coupling among the patches and the chaotic dynamics on the chaotic saddle. Specifically, the initial conditions of trajectories in all patches, or a subset of them, are set to approach the chaotic saddle instead of converging directly to the limit cycle on the $R$--$C$ plane with no predator species. Naturally, these trajectories are expected to remain in the vicinity of the chaotic saddle only for the aforementioned chaotic transients before escaping to the limit cycle. However, while the network trajectories are in the vicinity of the chaotic saddle, the dispersal, at a suitable dispersal rate, acts as a perturbation to the local dynamics of each patch. These perturbations disturb the patches' dynamics, causing re-injections of trajectories into the stable manifold of the chaotic saddle. As a consequence of such mutual perturbations among the patches, the trajectories remain in the vicinity of the chaotic saddle for an indefinitely long time interval, reinforcing the chaotic motion and, hence, causing the phenomenon of extended predator survival, as observed in Fig.~\ref{fig3}(b). This mechanism explains the observations reported in  Fig.~\ref{fig2}, Fig.~\ref{fig3}, and Fig.~\ref{fig4}. Specifically, if the dispersal rate is too low (left-hand corner of Fig.~\ref{fig4}), the random perturbations generated by the dispersal-coupling are not large enough to cause reinjections into the stable manifold of the chaotic saddle. As a consequence, the patches escape from the chaotic saddle at different times and reach the limit cycle on the $R$-$C$ plane at different phases, as observed in Fig.~\ref{fig3}(a). On the other hand, if the dispersal strength is too large (right-hand corner of Fig.~\ref{fig4}), the dispersal is sufficiently large to synchronize the patches' dynamics already in the vicinity of the chaotic saddle. As a consequence, the dispersal-coupling of the patches, which is proportional to the differences of the population densities, vanishes, and the trajectories of all patches escape the chaotic saddle unisonously, as observed in Fig.~\ref{fig3}(c). In this case, the network dynamics completely synchronize around the limit cycle on the $R$-$C$ plane.

Interestingly, this trapping mechanism has been previously described in the context of spatially extended physical systems \cite{Medeiros2018, Medeiros2019, medeiros2021impact}, where it has been demonstrated that escape times do not follow typical probability distributions. Despite the similarities, our linear and global stability analysis in Sec.~\ref{section4} reveals key differences in our ecological scenario. First, in our case, the synchronous configurations of the patches, corresponding to the extinction of predator species, remain unstable for most of the dispersal rate interval during which the trapping in the chaotic saddle occurs. By contrast, in the studied physical systems, the completely synchronized state is always stable and coexists with the asynchronous one. Second, in our ecological system, the trapping of trajectories in the chaotic saddle, which preserves predator species, also occurs for the globally coupled system. By contrast, in the studied physical systems, the trapping is suppressed for this coupling topology. We attribute these differences among the models to the specific interactions between the dispersal coupling and the local dynamics of the food chain.

\begin{figure*}[ht]
\centering
\includegraphics[width=0.6\linewidth]{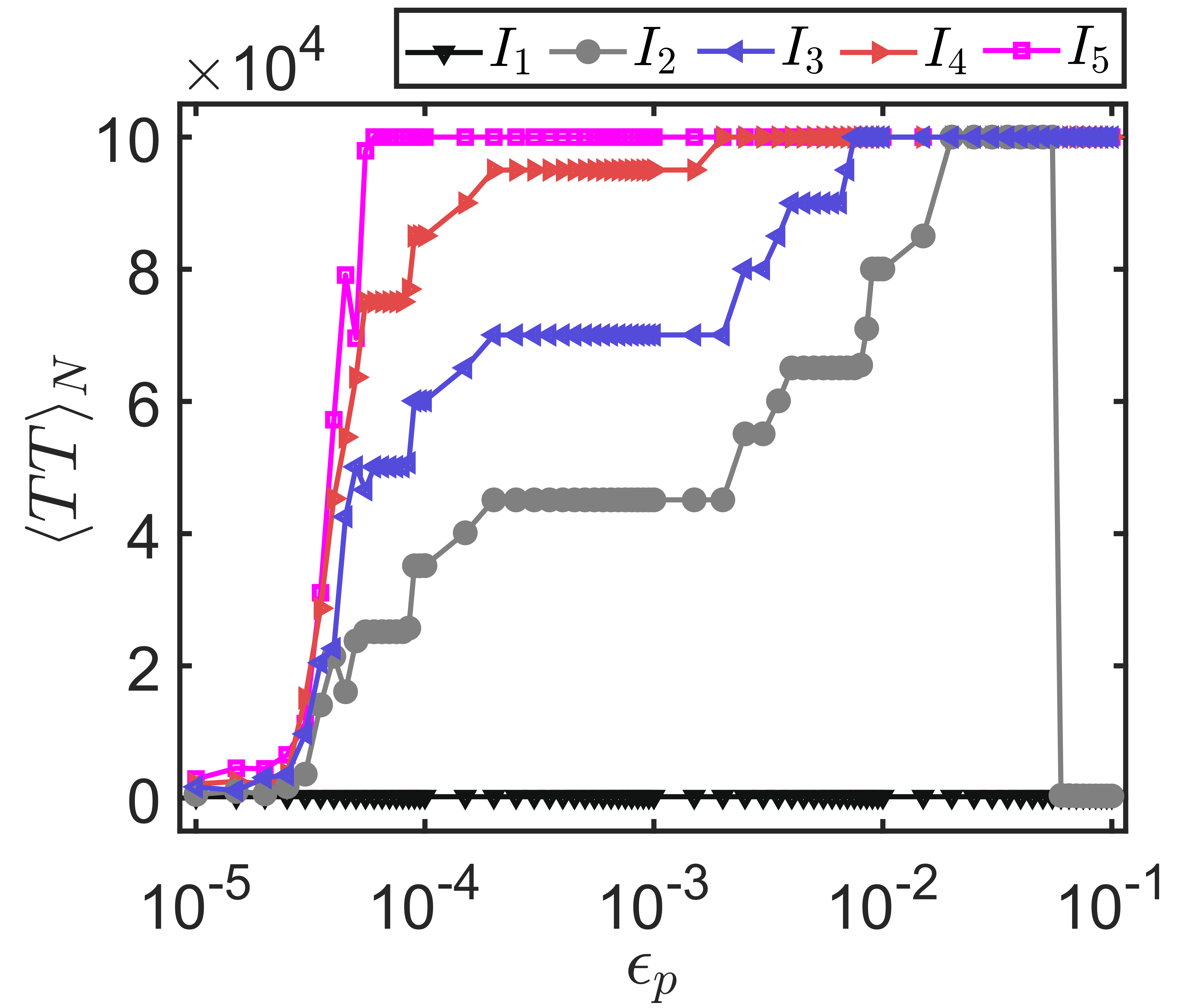}
\caption{Variation of the mean transient time for different choices of sets of initial conditions. Plots of $\langle TT \rangle_N$ versus $\epsilon_p \in [10^{-5}, 10^{-1}]$ are displayed for the five sets of collections of initial conditions (in brackets the relations between the fractions of initial conditions located in the former basin of attraction of the chaotic attractor and the $R$--$C$ limit cycle, respectively i.e., $I_1, 0\% : 100\%$ (black), $I_2, 25\% : 55\%$ (grey), $I_3, 50\% : 50\%$ (blue), $I_4, 75\% : 25\%$ (orange),$I_5, 100\% : 0\%$ (magenta), for the network topology $G_3$. $\epsilon_c$ is fixed at $10^{-5}$.}
\label{fig6} 
\end{figure*}

\begin{figure*}[ht]
\includegraphics[scale=0.5]{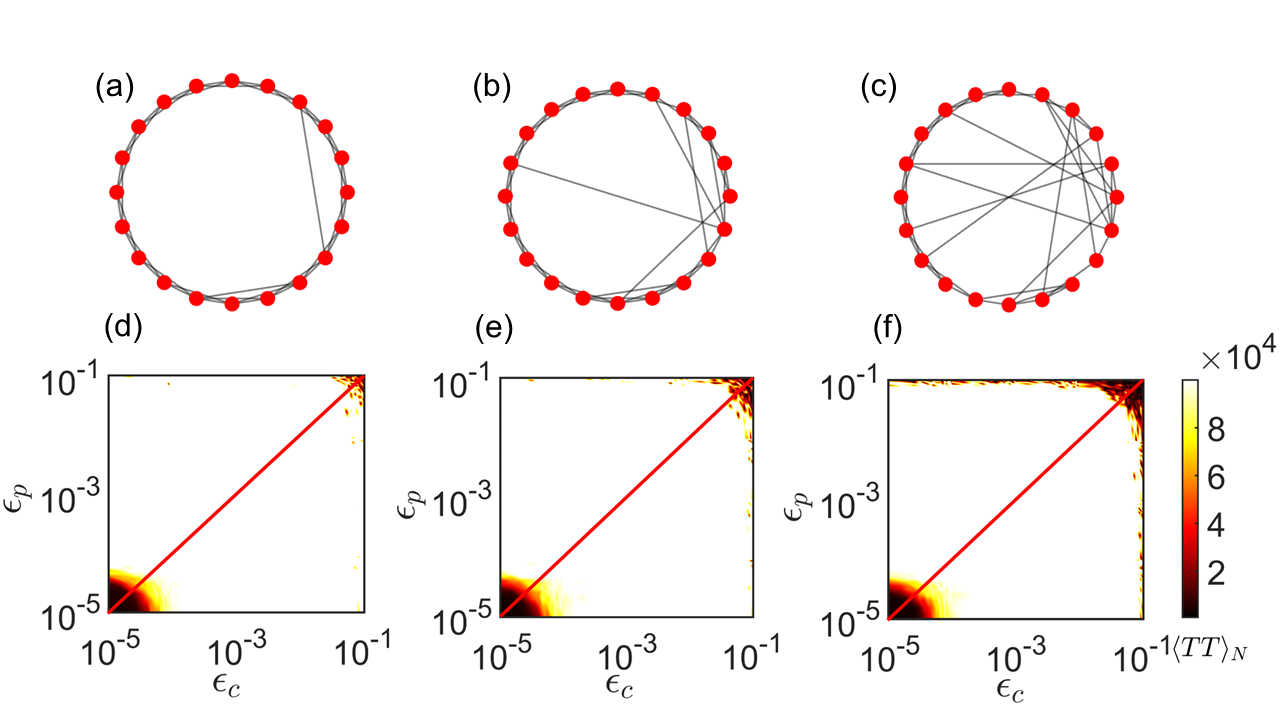}
\caption{(a)-(c) Network topology generated using Watts--Strogatz algorithm. We use the network (a) $G_3$ as a base network and perform the Watts-Strogatz algorithm on it to get two networks ((b) and (c)) with respective rewiring probabilities of $0.1$ and $0.2$. (d)-(f) $\epsilon_c-\epsilon_p$ parameter space for depicting $\langle TT\rangle_N$. The corresponding $\epsilon_c$-$\epsilon_p$ parameter spaces for the above-mentioned networks depicting $\langle TT \rangle_N$, respectively.}
\label{fig7} 
\end{figure*}

\section{Discussion}
\label{section6}
The structure of the asymmetric dispersal network emerges as a crucial determinant in predicting the destiny of ecological communities. The fact that groups of sub-populations can behave very differently depending on whether they are connected via dispersal or not is one of the most fundamental findings in spatial ecology \cite{abbott2011dispersal}. Metacommunities represented as ecological networks facing extinction often exhibit synchronous dynamics, which increases the risks of collapse \cite{earn2000coherence}. Against this backdrop, it is important to identify mechanisms that would prevent the loss of species by extinction. Here, we have demonstrated a possibility to prevent the extinction of species due to dispersal over a wide range of dispersal strengths. In the literature, the rescue of populations by dispersal among patches is well known in heterogeneous dispersal networks in which different patches possess different environmental conditions \cite{lehtinen2023empirical}. These rescue effects are solely dependent on this heterogeneity. Moreover, often only equilibrium states are taken into account. 

\par Here, we have unravelled a fundamentally different mechanism of rescue, which is characterized by a number of different factors  necessary for maintaining biodiversity due to asynchrony and asymmetry: 
\begin{enumerate}
\item The rescue mechanism is based on non-equilibrium solutions, emphasizing the importance of complex temporal dynamics for the maintenance of biodiversity. These chaotic dynamics ensure that species in each patch oscillate asynchronously. 
\item This mechanism can rescue predators even in homogeneous environments, where identical patch conditions typically promote synchronization and thus instability. A window in parameter space is identified in which, despite system identity, desynchronization emerges and enables species recovery. This is due to the occurrence of transient chaotic dynamics, which can be stabilized by dispersal over large intervals of dispersal strength. 

\item The topology of the dispersal networks matters as much as the dispersal strength. In particular, increasing the number of long-range connections progressively reduces the parameter region that supports prolonged transient chaos. 
Thus, not only link density but also the specific structural arrangement of dispersal pathways influences whether biodiversity can be maintained. Asymmetry in links as present in small world networks, best supports the survival of the predator. 
\item Initial conditions critically determine the persistence. When all patches are initialized in domains leading directly to extinction, dispersal alone cannot generate prolonged chaos. By contrast, if at least a subset of patches starts in a chaotic transient state, dispersal effectively revives and sustains chaotic oscillations across the network. This demonstrates that persistence requires the presence of transient chaotic trajectories in the initial configuration, which dispersal can then amplify and maintain.
\item Finally, it has been uncovered that the persistence of the predator does not require large measures to counteract the loss of the predator. Only a few corridors for dispersal between the patches are necessary to maintain biodiversity. While fully connected symmetric globally coupled networks are not beneficial, we found that sparse long-range connectivity, which is asymmetric, helps much more. It is demonstrated that small-world networks with a very small number of long-range connections are sufficient to rescue the predator, which is quite counterintuitive. The mechanism for this rescue of the predator relies on the dispersal coupling of the patches, which can be considered as a perturbation of the local dynamics, reinjecting it back into the chaotic saddle. This continuous reinjection sustains the chaotic oscillations indefinitely, allowing predator populations to persist rather than to collapse. 
\end{enumerate}
 
\par Harnessing such minimal connectivity offers a cost-effective and scalable principle for conservation. By strategically designing ecological corridors, restoring limited habitat linkages, vulnerable populations can be stabilized without the need for fully connected landscapes. These results suggest that even modest interventions in connectivity can unlock the stabilizing potential of transient chaos, turning what is usually considered a pathway to extinction into a mechanism of persistence. In fact, this study reveals the importance of the constructions of wildlife corridors in the past for nowadays problems of nature conservation. Such animal or wildlife crossings over highways and railways have been built already decades ago in many areas in the USA, Canada, and Europe (see e.g., Ontario Highway 69 Wildlife Crossing). Those crossings have been originally erected to prevent collisions between the traffic and animals crossing the road to save men and wildlife. Our study shows that those bridges and tunnels also serve another purpose, which the constructors did not have in mind many years ago and which is nowadays well recognized: they are also a very valuable contribution to maintaining biodiversity by helping predators to survive. More broadly, the findings underscore that biodiversity conservation in fragmented landscapes need not rely solely on extensive habitat restoration; rather, careful engineering of dispersal pathways may be sufficient to maintain ecological functioning. 

\section{Data Availability}
The computational frameworks used to derive the results described in this submission are openly available on the GitHub repository \cite{github}.

\section{acknowledgement}
D.G. was supported by the Science and Engineering Research Board (SERB), Government of India (Project No. CRG/2021/005894). E.S.M. and U.F. acknowledge support by the Deutsche Forschungsgemeinschaft (DFG) via Project No. 454054251 and by the São Paulo Research Foundation (FAPESP) via Project No. 2023/15040-0.

\appendix
\section*{APPENDIX-A: Effect of initial conditions}
\label{appendix1}
In the main text, we have chosen for our study initial conditions which are located in the basin of attraction of the chaotic attractor, even in cases where the boundary crisis has happened, and the chaotic attractor has turned into a chaotic saddle. In the latter case, the predator populations show transient chaos before going to extinction for the given choice of model parameters (see Sec. \ref{section2}). Besides the basin of attraction of the chaotic attractor there exists also the basin of attraction of the $R$--$C$ limit cycle and a random choice of an initial condition might affect the prolonged transient time of predator survival. To this end we now investigate the average transient time depending on the dispersal strength of the predator $\epsilon_p \in [10^{-5}, 10^{-1}]$ for the most effective case of network topology $G_3$. We choose 5 different scenarios depicted in Fig.~\ref{fig6}: the two extreme cases all initial conditions in the basins of the limit cycle $I_1$ and all in initial conditions in the former basin of attraction of the chaotic attractor $I_5$ are shown in black and magenta respectively, intermediate cases with different fractions of initial conditions from both basins of attraction are shown in grey, blue and orange $I_2, I_3, I_4$ respectively. As expected, if there were no initial conditions from the basin of attraction of the former chaotic attractor $I_1$, the average transient time is zero for all values of the dispersal strength, since all trajectory immediately converge to the limit cycle. For the other extreme case $I_5$ the average  transient time reaches quickly the longest simulation time. For a decreasing fraction of initial conditions exhibiting transient chaos it needs larger and larger dispersal strength to reach a longer and longer average. For the smallest nonzero fraction of initial conditions in the basin of the former chaotic attractor (25\%) we notice that for the largest coupling strength already synchronization occurs between the patches leading to the extinction of the predator. 

\section*{APPENDIX-B: Effect of rewiring in network topology}
\label{appendix2}
In Sec.~\ref{section3}, we have studied how the persistence of the predator population changes in the global network $G_1$ and the two random networks $G_2$ and $G_3$ that originated from an existing nonlocal network by the Newman--Watts small world algorithm. Here, we study how rewiring of links will affect the persistence of the predator population. We select the network $G_3$ in Fig.~\ref{fig7}(a) with link density $42$ and rewire the links (using the Watts--Strogatz algorithm) with probability $p=0.1$ and $p=0.2$ to construct the networks in Figs.~\ref{fig7}(b) and \ref{fig7}(c), respectively. We investigate how rewiring will affect the revival of prolonged transient chaos in our context. We compute the mean transient time $\langle TT \rangle_{N}$ of the three types of networks (Figs.~\ref{fig7}(d)-(f)) by simultaneously changing $\epsilon_c \in [10^{-5},10^{-1}]$ and $\epsilon_p \in [10^{-5},10^{-1}]$. We observe that the more we increase the rewiring probability, the more the white portion in each figure decreases, resulting in a diminishing  range of coupling strength allowing for the persistence of prolonged transient chaos which supports the survival of the predator.

\bibliography{reference_arx}

\end{document}